\documentclass[9pt,shortpaper,twoside,web]{ieeecolor}
\usepackage{generic}
\usepackage{cite}
\usepackage{amsmath,amssymb,amsfonts}
\usepackage{algorithmic}
\usepackage{graphicx}
\usepackage{subcaption}
\usepackage[font={small,it}]{caption}
\usepackage{textcomp}
\def\BibTeX{{\rm B\kern-.05em{\sc i\kern-.025em b}\kern-.08em
    T\kern-.1667em\lower.7ex\hbox{E}\kern-.125emX}}
\begin{document}
\title{Entanglement control of two-level atoms in dissipative cavities}
\author{Reyad Salah, Ahmed M. Farouk, Ahmed Farouk, M. Abdel-Aty, and A.-S. F. Obada
\thanks{\textbf{Reyad Salah} (\emph{reyadkhalf@gmail.com}) is with Department of Mathematics, Faculty of Science, Helwan University, Helwan, Egypt.}
\thanks{\textbf{Ahmed M. Farouk} (\emph{ahmed.farouk@azhar.edu.eg}) and \textbf{A.-S. F. Obada} (\emph{asobada@yahoo.com}) are with Department of Mathematics, Faculty of Science, Al-Azhar University, Nasr City 11884, Cairo, Egypt.}
\thanks{\textbf{Ahmed Farouk} (\emph{dr.ahmedfarouk85@yahoo.com}) is with Department of Physics and Computer Science, Wilfrid Laurier University,	Waterloo, Canada.} 
\thanks{\textbf{M. Abdel-Aty} (\emph{mabdelaty@zewailcity.edu.eg})is with Zewail University of Science and Technology, Zayed City 12588, Giza, Egypt.}}

\maketitle

\begin{abstract}
An open quantum bipartite system consisting of two independent two-level atoms interacting non-linearly with a two-mode electromagnetic cavity field is investigated by proposing a suitable non-Hermitian generalization of Hamiltonian. The mathematical procedure of obtaining the corresponding wave function of the system is clearly given. Panchartnam phase is studied to give a precise information about the required initial system state, which is related to artificial phase jumps, to control the Degree of Entanglement (DEM) and get the highest Concurrence. We discuss the effect of time-variation coupling, and dissipation of both atoms and cavity. The effect of the time-variation function appears as frequency modulation (FM) effect in the radio waves. Concurrence rapidly reaches the disentangled state (death of entanglement) by increasing
the effect of field decay. On the contrary, the atomic decay has no effect.
\end{abstract}


\section{Introduction}
\label{sec:introduction}
Quantum systems promise enhanced capabilities in sensing, communications and computing beyond what can be achieved with classical-based conventional technologies rather than quantum physics. Mathematical models are essential for analyzing these systems and building suitable quantum models from empirical data is an important research topic. In Dirac theory of radiation \cite{louisell1973quantum}, he considered atoms and the radiation field with which they interact as a single system whose energy is represented by the frequency/energy of each atom solely, the frequency/energy of every mode of the applied laser field alone and a small term is to the coupling energy between atoms and field modes. The interaction term is necessary if atoms and field modes are to affect each other. A simple model is that we consider a pendulum of resonant frequency $\omega_0$, which corresponds to an atom, and a vibrating string of resonant frequency $\omega_1$ which corresponds to the radiation field. Jaynes-Cummings model (JCM) \cite{Jayness1963} is the first solvable analytical model to represent the atom-field interaction with experimental verification \cite{rempe1987observation}. JCM has been subjected to intensive research in the last decades with many interesting phenomena  explored \cite{eberly1980periodic, meystre1982squeezed, shore1993jaynes, mahran1989amplitude}. The matter-field coupling term may be constant \cite{abdel1987n,abdalla1990dynamics,abdel2003shannon} or time-dependent \cite{buvzek1989jaynes,hood1998real} and that depends on the considered physical situation. In our case, the atoms are moving while interacting with the laser field, this topic has been investigated for different quantum systems\cite{babiker1994light,prants1997dynamical,abdel2008effect,abdel2010geometric,eleuch2012effects}

Using a Non-Hermitian generalization of Hamiltonian (NHH) is now considered as a model to describe an open quantum system \cite{eleuch2014open,eleuch2015nearby,eleuch2017resonances}, we may obtain complex-energy eigenvalues. These NHHs are justified as an approximate and phenomenological description of an open quantum system such as radioactive decay processes\cite{bender2007making}. Driving a quantum system with the output field from another driven quantum system, and a quantum trajectory theory for cascaded open systems were studied by proposing NHH in \cite{gardiner1993driving}, and \cite{carmichael1993quantum}, respectively. Investigating the dynamics of three-level systems has allowed the discussion of teleportation and non-classical properties \cite{daneshmand2016effects,lee2000transfer}, Concurrence and Shannon information entropy \cite{obada2017moving,ismail2019generation}.

In this work, we propose a new technique to control the entanglement. Stimulated Raman adiabatic passage (STIRAP) is a process that allows transfer of a population between two states via at least two coherent field pulses by inversely engineering the Hamiltonian parameters via Lewis-Riesenfeld phases \cite{torrontegui2013shortcuts}. STIRAP has been explained chemically and physically \cite{vitanov2017stimulated} and its protocols have been applied to various models; two-level atom \cite{chen2011lewis}, three-level atom \cite{giannelli2014three,chen2012engineering}, and four-level atom\cite{issoufa2015generation}. The choice of initial system parameters as we propose is related to the artificial phase jumps of Pancharatnam phase. Phase jumps are promising points such that it generates better entanglement degrees and its successive repetition inside any system dynamics reflects a good sign of system capability to transfer information, as the geometric phases can be altered by changing the relative delay of the laser pulses \cite{unanyan1999laser}. 

Our work here is oriented around  the interaction of an open quantum system of two independent two-level atoms with a quantization (non-classical) of electromagnetic field in a dissipative cavity in the multi-photon process. In section \ref{section-model}, the considered physical scenario is introduced, the corresponding Hamiltonian is investigated and the mathematical procedure for obtaining the solution of the wave equation is clearly given. In section \ref{section-panch}, we discuss the proposed technique to control the entanglement by properly choosing the initial values of the atomic state. Concurrence is also discussed to determine the effect of other parameters in the system. In section \ref{section-conclusion}, a brief conclusion and results are given.

\section{Physical scenario}\label{section-model}
\begin{figure}[!htb]
	\centering
	\includegraphics[width=0.8\columnwidth,height=3.5cm]{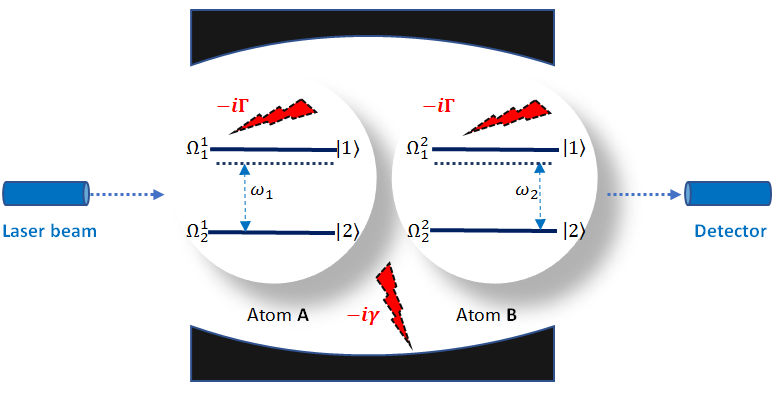}
	\caption{Energy level diagram of two two-level atoms coupled to two-mode field in a dissipative cavity.}
	\label{model-figure}
\end{figure}
The theoretical model as illustrated in figure(\ref{model-figure}) can be written as a non-Hermitian Hamiltonian
\begin{equation} \label{Hamiltonian}
\begin{split}
\hat{H}=\frac{1}{2}\sum_{j=1}^{2}\sum_{\ell=1}^{2}&(-1)^{j+1}(\Omega_{j}^{\ell}-i\Gamma_{j}^{\ell})\hat{\sigma}_{jj}^{\ell}+\sum_{j=1}^{2}(\omega_{j}-i\gamma_{j}) \hat{a}_{j}^{\dagger}\hat{a}_{j}\\&+\sum_{\ell=1}^{2}\bar{\lambda}_{\ell}(t) \bigl(\hat{a}_{\ell}^{\kappa} \hat{\sigma}_{12}^{\ell} + \hat{a}_{\ell}^{\dagger \kappa} \hat{\sigma}_{21}^{\ell} \bigl).
\end{split}
\end{equation}
where $\Omega_{j}^{\ell}$ is the associated frequency of level $j$ of the corresponding atom $\ell$, with $\Gamma_{j}^{\ell}$ is the atomic corresponding decay rate, and $\hat{\sigma}_{ij}^{\ell}=\left|i\right\rangle \left\langle j\right|$, ($i,j=1,2$) are the atomic-flip operators for $\left|j\right\rangle\rightarrow\left|i\right\rangle$, they satisfy the commutation relation $[\hat{\sigma}_{ij},\hat{\sigma}_{\alpha \beta}]=\hat{\sigma}_{i\beta}\delta_{\alpha j}-\hat{\sigma}_{\alpha j}\delta_{i\beta}$. $\omega_{j}$ is the frequency of the quantized electromagnetic cavity field mode $j$ with a corresponding decay rate $\gamma_{j}$ and $\hat{a}_{j}$ ($\hat{a}_{j}^{\dagger}$) is the annihilation (creation) operator for the field mode $j$, and they obey the commutation relation $[\hat{a}_{i},\hat{a}_{j}^{\dagger}]=\delta_{ij}$. Here, we consider that $\Omega_{j}^{\ell}>>\Gamma_{j}^{\ell}$, and $\omega_{j}^{\ell}>>\gamma_{j}^{\ell}$ \cite{bhattacherjee2018realization}. $\bar{\lambda}_{\ell}(t)$ is the time-dependent coupling  of the matter-field interaction. It is more realistic to consider that the interaction intensity is not uniform, in the following calculations we consider that $\bar{\lambda}_{\ell}(t)=\lambda_{\ell}\cos(\varpi_{\ell}t)$. To study the dynamics and properties of this model, we need to get the corresponding wave function $\left|\psi(t)\right\rangle$, which can be formulated in the following form;
\begin{equation}
\begin{split}
\left|\psi(\phi,t)\right\rangle=\sum_{n_1,n_2=0}^{\infty}&\bigg( A^{(n_1,n_2)}_{1} e^{-i\alpha_{1} t} \left|n_1,n_2,1,1\right\rangle \\&\quad + A^{(n_1,n_2)}_{2} e^{-i\alpha_{1} t} \left|n_1,n_2+\kappa,1,2\right\rangle \\&\quad + A^{(n_1,n_2)}_{3}(\phi,t) e^{-i\alpha_{1} t} \left|n_1+\kappa,n_2,2,1\right\rangle \\&\quad+ A^{(n_1,n_2)}_{4}e^{-i\alpha_{1} t} \left|n_1+\kappa,n_2+\kappa,2,2\right\rangle\bigg)
\end{split}
\end{equation}
where $A_{m}^{(n_1,n_2)}(t)$ ($m=1,2,3,4$) are functions of time and field modes, called the probability amplitudes. $\alpha_{m}$ are field-dependent functions, and can be defined as follows;
\begin{equation}
\begin{split}
&\alpha_{1}=\frac{1}{2}(\Omega_{1}^{1}-\Omega_{1}^{2})+n_1\omega_1+n_2\omega_2,\\&
\alpha_{2}=\frac{1}{2}(\Omega_{1}^{1}-\Omega_{2}^{2})+n_1\omega_1+(n_2+\kappa)\omega_2,\\&
\alpha_{3}=\frac{1}{2}(\Omega_{1}^{2}-\Omega_{2}^{1})+(n_1+\kappa)\omega_1+n_2\omega_2,\\&
\alpha_{4}=\frac{-1}{2}(\Omega_{2}^{1}+\Omega_{2}^{2})+(n_1+\kappa)\omega_1+(n_2+\kappa)\omega_2.
\end{split}
\end{equation}
By applying the time-dependent Schr\"{o}dinger equation to the system, we get the following coupled differential equations. The trigonometric function in $\bar{\lambda}_{\ell}(t)$ can be reformulated in an exponential form. There exist exponential terms with two different powers in the differential equations, $e^{\pm i(\Delta+\varpi)t}$ and $e^{\pm i(\Delta-\varpi)t}$. Approximately, we can ignore the counter oscillating terms $e^{\pm i(\Delta+\varpi)t}$. This approximation is similar to the RWA which is used in plethora of physical models \cite{obada2017moving,louisell1961quantum};
\begin{equation} \label{coupled-system}
\begin{split}
& i \left[ 
\begin{array}{cccc}
\dot{A}_1^{(n_1,n_2)}(t) & \dot{A}_2^{(n_1,n_2)}(t) & \dot{A}_3^{(n_1,n_2)}(t) & \dot{A}_4^{(n_1,n_2)}(t)
\end{array}
\right]^{T} \\&=\left[ \begin{array}{cccc}
k_1&g_2 e^{i\delta t} & g_1 e^{i\delta t} &0 \\
g_2 e^{-i\delta t} & k_2 & 0 & g_1 e^{i\delta t} \\
g_1 e^{-i\delta t} & 0 & k_3 & g_2 e^{i\delta t} \\
0 & g_2 e^{-i\delta t} & g_1 e^{-i\delta t} & k_4
\end{array}\right]\times\\&\times
\left[ 
\begin{array}{cccc}
A_1^{(n_1,n_2)}(t) & A_2^{(n_1,n_2)}(t) & A_3^{(n_1,n_2)}(t) & A_4^{(n_1,n_2)}(t)
\end{array}
\right]^{T}
\end{split}
\end{equation} 
where
$$ \delta=\Delta-\varpi,$$
$$k_1=\frac{1}{2i}\left(\Gamma_1^{1}+\Gamma_1^{2}+n_1\gamma_{1}+n_2\Gamma_{2}\right),$$

$$ k_2=\frac{1}{2i}\left(\Gamma_1^{1}-\Gamma_2^{2}+n_1\gamma_{1}+(n_2+\kappa)\Gamma_{2}\right), $$

$$k_3=\frac{1}{2i}\left(\Gamma_1^{2}+\Gamma_2^{1}+(n_1+\kappa)\gamma_{1}+n_2\Gamma_{2}\right),$$

$$k_4=\frac{1}{2i}\left(-\Gamma_2^{1}-\Gamma_2^{2}+(n_1+\kappa)\gamma_{1}+(n_2+\kappa)\Gamma_{2}\right),$$
$$ \Delta=\Omega_{j}^{1}-\kappa \omega_1=\Omega_{j}^{2}-\kappa \omega_2,$$
$$
g_{1}=\frac{\bar{\lambda}_1}{2} \sqrt{\dfrac{(n_1+\kappa)! }{n_1!}},$$

$$g_{2}=\frac{\bar{\lambda}_2}{2} \sqrt{\dfrac{(n_2+\kappa)! }{n_2 !}}.$$
After using the following transformation 
\begin{align*} \label{gamma1}
&B_1^{(n_1,n_2)}(t)= A_1^{(n_1,n_2)}(t) e^{i(\Delta-\varpi)t},\\&
B_2^{(n_1,n_2)}(t)= A_2^{(n_1,n_2)}(t),\\&
B_3^{(n_1,n_2)}(t)= A_3^{(n_1,n_2)}(t),\\&
B_4^{(n_1,n_2)}(t)= A_4^{(n_1,n_2)}(t) e^{-i(\Delta-\varpi)t},
\end{align*}
we get
\begin{equation} \label{eq-3}
i \left[
\begin{array}{cccc}
\dot{B}_1^{(n_1,n_2)}(t) \\ \dot{B}_2^{(n_1,n_2)}(t) \\ \dot{B}_3^{(n_1,n_2)}(t) \\ \dot{B}_4^{(n_1,n_2)}(t)
\end{array}
\right]=\left[
\begin{array}{cccc}k_1&g_2  & g_1 &0 \\
g_2 & k_2 & 0 & g_1 \\
g_1 & 0 & k_3 & g_2 \\
0 & g_2  & g_1  & k_4\end{array}
\right]  \left[
\begin{array}{ccc}
B_1^{(n_1,n_2)}(t) \\ B_2^{(n_1,n_2)}(t) \\  B_3^{(n_1,n_2)}(t) \\  B_4^{(n_1,n_2)}(t)
\end{array}
\right]
\end{equation}

This coupled system of differential equations can be solved analytically. The energy eigenvalues $\mathcal{E}_m(t)$ of the system in equation (\ref{Hamiltonian}), can be formulated as follows;
\begin{equation}
\mathcal{E}_m(t)=-\frac{b}{4}\pm s\pm\sqrt{\frac{q}{s}-2p-4s^2},
\end{equation}
with
\begin{equation}
\begin{split}
&s=\frac{1}{2}\sqrt{\frac{1}{3}\left(\mathcal{Q}+\frac{\Delta_{0}}{\mathcal{Q}}\right)},\\&
\mathcal{Q}=\left( \frac{\Delta_1+\sqrt{\Delta_1^{2}-4\Delta_{0}^{3}}}{2} \right)^{1/3},\\& 
p=\frac{8c-3b^2}{8},\\& q=\frac{b^3+8d-4bc}{8}, \\& \Delta_0=c^2-3bd+12e, \\& \Delta_1=2c^3-9bcd+27(b^2 e + d^2)-72 c e, 
\end{split}
\end{equation}
\begin{equation}
\begin{split}
&
b=-i(\kappa_1+\kappa_2+\kappa_3+\kappa_4),
\\&
c=\Delta^{2}+2g_{1}^{2}+2g_{2}^{2}+\Delta \kappa_1-\kappa_1\kappa_2-\kappa_1\kappa_3-\kappa_2\kappa_3-\Delta \kappa_4 \\& \quad -\kappa_1\kappa_4-\kappa_2\kappa_4-\kappa_3\kappa_4-2 \Delta \varpi-\varpi\kappa_1+\varpi \kappa_4+\varpi^{2},
\\&
d=-i\bigg(\Delta \kappa_2+\Delta \kappa_1\kappa_2+\Delta^{2}\kappa_3+\Delta \kappa_1\kappa_3-\kappa_1\kappa_2\kappa_3-\Delta\kappa_2\kappa_4\\&\quad-\kappa_1\kappa_2\kappa_4 - \Delta\kappa_3\kappa_4-\kappa_1\kappa_3\kappa_4-\kappa_2\kappa_3\kappa_4+g_1^{2}(\kappa_1+\kappa_2\\&\quad+\kappa_3+\kappa_4)+ g_2^{2}(\kappa_1+\kappa_2+\kappa_3+\kappa_4)-(\kappa_2+\kappa_3)\\&\quad\times(2\Delta+\kappa_1-\kappa_4)\varpi+\varpi^{2}(\kappa_2+\kappa_3)\bigg),
\\&
e = g_1^{4}+\bigg(g_2^{2}-\kappa_2(\Delta+\kappa_1-\varpi)\bigg) \bigg(g_2^{2}+\kappa_3(\Delta-\kappa_4-\varpi)\bigg) \\& \quad  - g_1^{2}\bigg(2g_2^{2} + \Delta(\kappa_3-\kappa_2) + \kappa_3(\kappa_1-\varpi)+\kappa_2(\kappa_4+\varpi)\bigg).
\end{split}
\end{equation}
%%%%%%%%%%%%%%%%%%%%%%%%%%%%%%%%%%%%%%%%%%%%%%%%%%%%%%%%%%%%
%\section{Derivation of the exponential of the matrix}
\quad By applying \textit{Newton interpolation} method \cite{moler1978nineteen} for getting the  matrix exponential, which states that for a matrix $A$ with eigenvalues $\lambda_j$, ($j=1,2,..,n$), $n$ is the dimension of the matrix,
\begin{equation}
e^{tA}=e^{\lambda_1 t}\mathcal{I}+\sum_{j=2}^{n}\left[ \lambda_1,...,\lambda_j \right]\Pi_{\kappa=1}^{j-1}\left(A-\lambda_{\kappa}\mathcal{I}\right),
\end{equation}
where $\mathcal{I}$ is the unitary matrix and the divided differences $\left[ \lambda_1,...,\lambda_j \right]$ depend on $t$ and defined recursively by:
\begin{equation}
\left[ \lambda_1,\lambda_2 \right]=\frac{e^{\lambda_1 t}-e^{\lambda_2 t}}{\lambda_1-\lambda_2}
\end{equation}
\begin{equation}
\left[ \lambda_1,...,\lambda_{\kappa+1} \right]=\frac{\left[ \lambda_1,...,\lambda_{\kappa} \right]-\left[ \lambda_2,...,\lambda_{\kappa} \right]}{\lambda_1-\lambda_{\kappa+1}}, \qquad \kappa\geq 2.
\end{equation}
So by using the above method to $e^{-i\mathcal{M}t}$ where $A=-i\mathcal{M}$ and the eigenvalues of $A$ are defined in eq.(\ref{eq-3}). Then
\begin{equation}
\begin{split}
e^{-it\mathcal{M}}&=e^{\mathcal{E}_1 t}\mathcal{I}+\left[\mathcal{E}_1,\mathcal{E}_2\right]\left(-i\mathcal{M}-\mathcal{E}_1\mathcal{I}\right)\\&\quad+\left[\mathcal{E}_1,\mathcal{E}_3\right]\left(-i\mathcal{M}-\mathcal{E}_1\mathcal{I}\right)\left(-i\mathcal{M}-\mathcal{E}_2\mathcal{I}\right)\\&\quad+\left[\mathcal{E}_1,\mathcal{E}_4\right]\left(-i\mathcal{M}-\mathcal{E}_1\mathcal{I}\right)\left(-i\mathcal{M}-\mathcal{E}_2\mathcal{I}\right)\left(-i\mathcal{M}-\mathcal{E}_3\mathcal{I}\right)\\&\quad+\left[\mathcal{E}_1,\mathcal{E}_5\right]\left(-i\mathcal{M}-\mathcal{E}_1\mathcal{I}\right)\left(-i\mathcal{M}-\mathcal{E}_2\mathcal{I}\right)\\&\qquad\times\left(-i\mathcal{M}-\mathcal{E}_3\mathcal{I}\right)\left(-i\mathcal{M}-\mathcal{E}_4\mathcal{I}\right),
\end{split}
\end{equation}
where the divided differences are formulated as:
\begin{equation}\label{divided-1}
\left[\mathcal{E}_1,\mathcal{E}_2\right]=\frac{e^{\mathcal{E}_1 t}-e^{\mathcal{E}_2 t}}{\mathcal{E}_1-\mathcal{E}_2},
\qquad\qquad\qquad\qquad\qquad\qquad
\end{equation}
\begin{equation}
\begin{split}
\left[\mathcal{E}_1,\mathcal{E}_3\right]&=\frac{\left[\mathcal{E}_1,\mathcal{E}_2\right]-\left[\mathcal{E}_2,\mathcal{E}_3\right]}{\mathcal{E}_1-\mathcal{E}_3}\\&=\frac{e^{\mathcal{E}_1 t}-e^{\mathcal{E}_2 t}}{(\mathcal{E}_1-\mathcal{E}_2)(\mathcal{E}_1-\mathcal{E}_3)}-\frac{e^{\mathcal{E}_2 t}-e^{\mathcal{E}_3 t}}{(\mathcal{E}_1-\mathcal{E}_2)(\mathcal{E}_2-\mathcal{E}_3)},
\end{split}
\end{equation}
\begin{equation}
\begin{split}
\left[\mathcal{E}_1,\mathcal{E}_4\right]&=\frac{\left[\mathcal{E}_1,\mathcal{E}_3\right]-\left[\mathcal{E}_3,\mathcal{E}_4\right]}{\mathcal{E}_1-\mathcal{E}_4}\\&=\frac{e^{\mathcal{E}_1 t}-e^{\mathcal{E}_2 t}}{(\mathcal{E}_1-\mathcal{E}_2)(\mathcal{E}_1-\mathcal{E}_3)(\mathcal{E}_1-\mathcal{E}_4)}\\&\quad-\frac{e^{\mathcal{E}_2 t}-e^{\mathcal{E}_3 t}}{(\mathcal{E}_1-\mathcal{E}_2)(\mathcal{E}_2-\mathcal{E}_3)(\mathcal{E}_1-\mathcal{E}_4)}-\frac{e^{\mathcal{E}_3t}-e^{\mathcal{E}_4 t}}{(\mathcal{E}_1-\mathcal{E}_4)(\mathcal{E}_3-\mathcal{E}_4)}.
\end{split}
\end{equation}

%%%%%%%%%%%%%%%%%%%%%%%%%%%%%%%%%%%%%%%%%%%%%%%%%%%%%%%%%%%%%%%%%%%%%%%%%%%%%%%%%%%%%%%%%%%%%%%%%%%%%%%%%%%%%%%%%%%%%%%%%%%%%%%%%%%%%%%%%%%%%%%%%%%%%%%%%%%%%%%%%%%%%%%%%%%%%%
\quad After the derivation of the exponential of the matrix, we can calculate the formulas of the probability amplitudes of the wave function of the sytem. The atoms are initially in superposition of states i.e.  $\left|\psi(\phi,0)\right\rangle_{A}= \cos(\theta)\left|1,1\right\rangle + e^{-i\phi} \sin(\theta)\left|2,2\right\rangle$, and the initial field is oriented in the coherent states. Then the final form of the probability amplitudes are
\begin{equation}
\left[
\begin{array}{ccc}
A_1^{(n_1,n_2)}(t) e^{i(\Delta-\varpi)t}\\ A_2^{(n_1,n_2)}(t) \\  A_3^{(n_1,n_2)}(t) \\  A_4^{(n_1,n_2)}(t) e^{-i(\Delta-\varpi)t}
\end{array} \right]= e^{-it\mathcal{M}} \left[
\begin{array}{ccc}
A_1^{(n_1,n_2)}(0) \\ A_2^{(n_1,n_2)}(0) \\  A_3^{(n_1,n_2)}(0) \\  A_4^{(n_1,n_2)}(0)
\end{array}
\right]
\end{equation}
%%%%%%%%%%%%%%%%%%%%%%%%%%%%%%%%%%%%%%%%%%%%%%%%%%%%%%%%%%%%%%%%%%%%

%%%%%%%%%%%%%%%%%%%%%%%%%%%%%%%%%%%%%%%%%%%%%%%%%%%%%%%%%%%%
\begin{figure}[b]
	\centering
	\includegraphics[width=0.8\columnwidth,height=3.5cm]{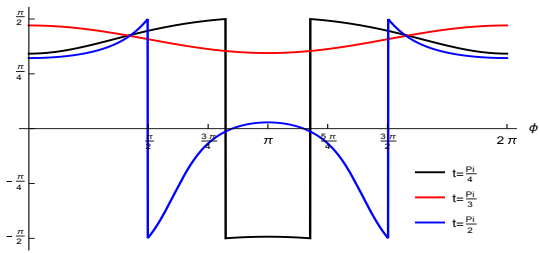}
	\caption{The evolution of Pancharatnam phase $\varphi(\phi)$}\label{control-phase}
\end{figure}
\section{Pancharatnam phase and Concurrence}\label{section-panch}
\quad We need to estimate a certain parameter for controlling the dynamics and entanglement of the system. A special attention is paid for the value of the initial latter phase parameter $\phi$. To reach that goal, we investigate the evolution of Pancharatnam phase $\Phi(\phi,t)=\arg(\left\langle\psi(\phi,0)|\psi(\phi,t)\right\rangle)$. To control the phase $\phi$, we plot $\Phi(\phi,t)$ vs $\phi$ for three different values of the scaled time $\lambda t$, as in figure (\ref{control-phase}). The red, black, and blue curves are plotted for $\lambda t=\pi/3$, $\pi/4$, and $\pi/2$, respectively. In the red curve, we note that there is a smooth evolution of the phase, while for the black and blue curves, they exhibit two artificial phase jumps for two different values of $\phi$. The phase jump for the blue curve ($\lambda t=\pi/2$) is repeated every period of $\pi$ and in-between the jumps the evolution is semi-parabolic shaped and reflects a slow variation of the system. The two phase jumps of the black curve are repeated every $\approx 13\pi/20$ and in-between the two jumps the variation is very slow, smooth and is separated by $\approx 7\pi/20$.

\begin{figure}[!htb]\centering
	\begin{subfigure}[b]{0.45\columnwidth}
		\centering
		\includegraphics[width=\textwidth,height=3cm]{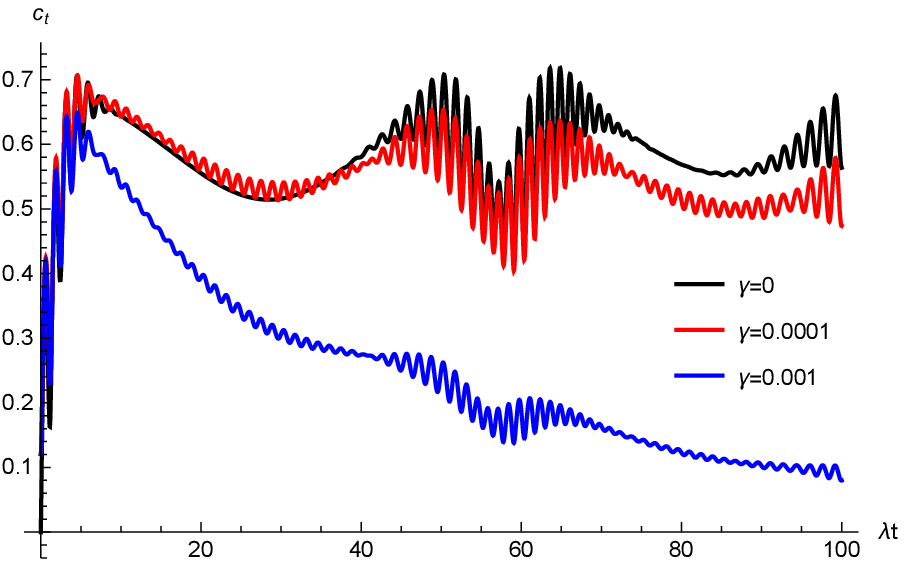}
		\caption{$\varpi=\pi$, $\theta=0$.}\label{conc3}
	\end{subfigure}
	\quad
	\begin{subfigure}[b]{0.45\columnwidth}
		\centering
		\includegraphics[width=\textwidth,height=3cm]{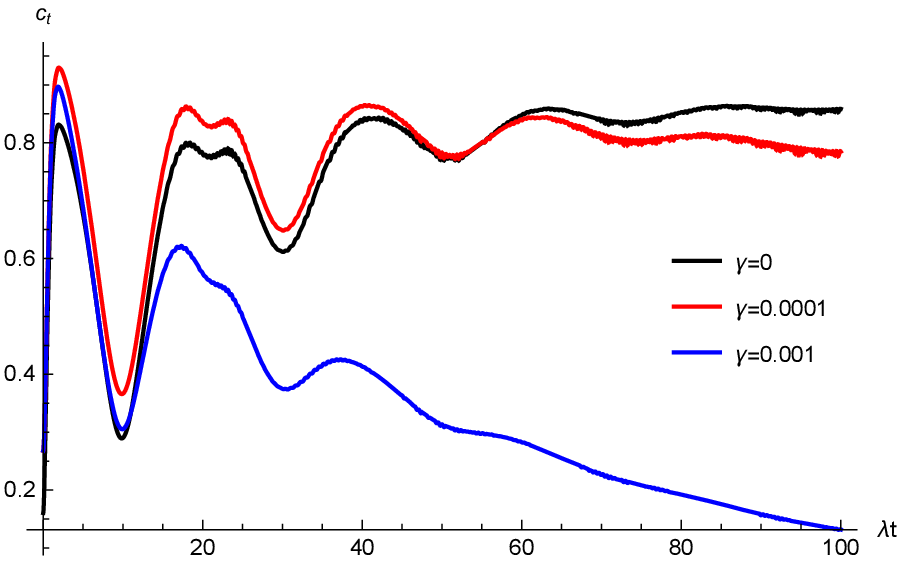}
		\caption{$\varpi=0$, $\theta=\pi/4$.}\label{conc1}
	\end{subfigure}
	\quad
	\begin{subfigure}[b]{0.45\columnwidth}
		\centering
		\includegraphics[width=\textwidth,height=3cm]{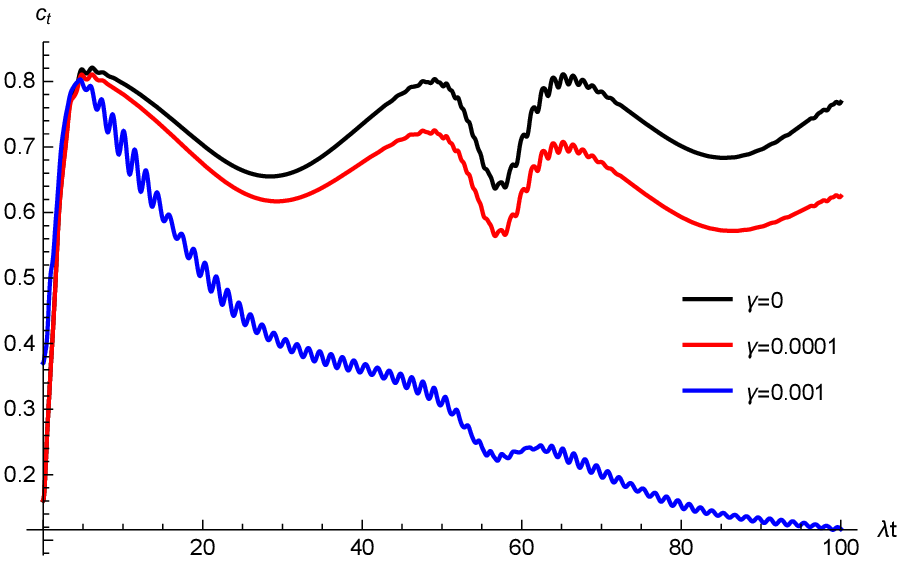}
		\caption{$\varpi=\pi$, $\theta=\pi/4$.} \label{conc2}
	\end{subfigure}
	\caption{The evolution of \textit{concurrence} $\mathcal{C}(t)$ vs the scaled time, for the one-photon process ($\kappa=1$), $\omega_{j}=\Omega_{j}=0.1\lambda$, $\Gamma_{j}=0$, and $\bar{n}_{j}=10$.}
	\label{concurrence-figures}
\end{figure}

Now, we can detect the dynamical behavior of the considered mutipartite system, by investigating the Degree of Entanglement (DEM) by using the \textit{concurrence} measure, which was formulated as a convex measure to amount the DEM for two qubits in pure states by Wootters and Hills \cite{hill1997entanglement}. For two qubits in pure states, concurrence is $\mathcal{C}(t)=\sqrt{2\left( 1-Tr \hat{\varrho}^2 \right)}$, where $\hat{\varrho}=\left|\psi(t)\right\rangle\left\langle\psi(t)\right|$ is the reduced density operator. The definition of concurrence has been extended to include multiple qubits \cite{abdel2007quantum}, and can be calculated generally by;
\begin{equation*}
\mathcal{C}(t)=\sqrt{2\sum_{i,j=1, i\neq j}^{4}\left( \varrho_{ii}\rho_{jj}-\varrho_{ij}\varrho_{ji}\right)},
\end{equation*}
where $\varrho_{ij}$ are the elements of reduced density $\varrho$ in matrix form.  Figures (\ref{concurrence-figures}) sketches the evolution of \textit{concurrence} $\mathcal{C}(t)$ against the scaled time $\lambda t$.

\begin{figure}[!htb]
	\begin{subfigure}[b]{0.45\columnwidth}
		\centering
		\includegraphics[width=\textwidth,height=3cm]{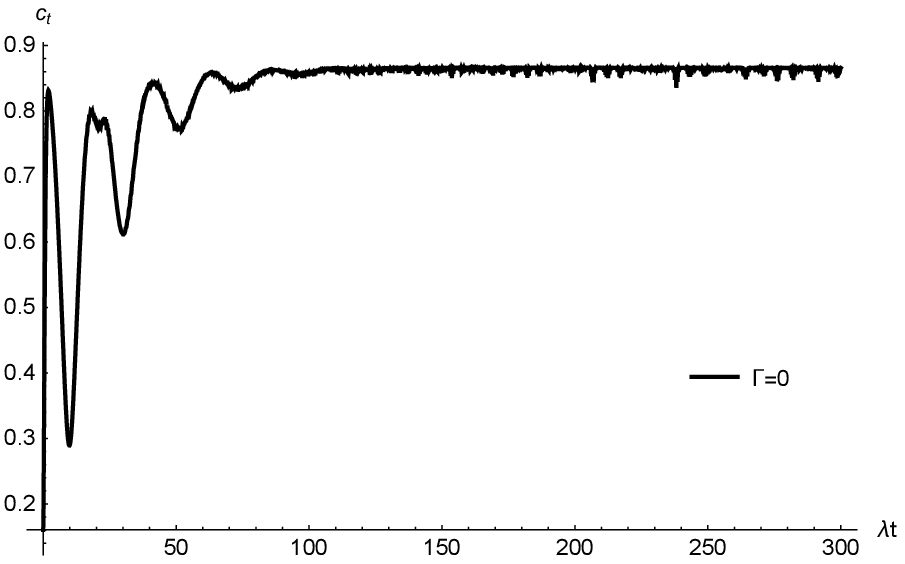}
		\caption{$\varpi=0$}\label{1atom}
	\end{subfigure}
	\quad
	\begin{subfigure}[b]{0.45\columnwidth}
		\centering
		\includegraphics[width=\textwidth,height=3cm]{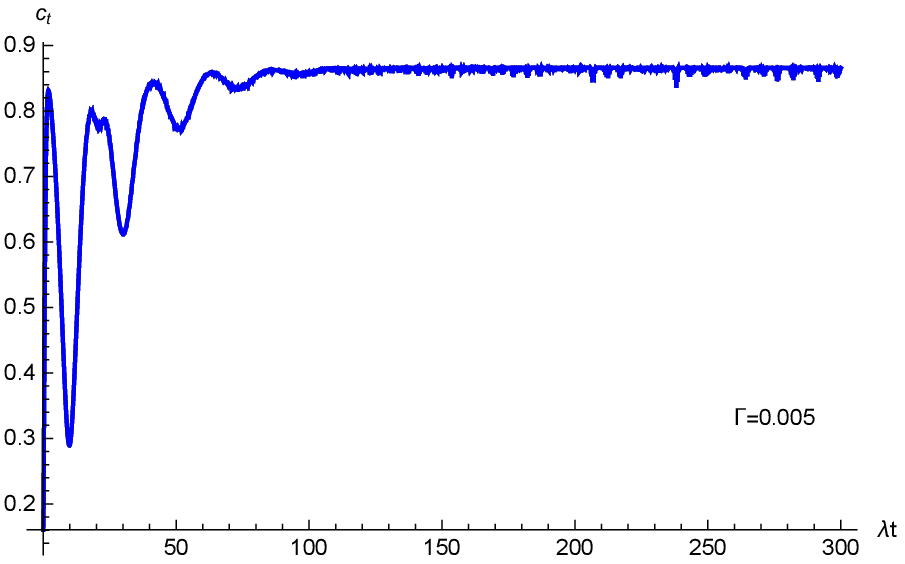}
		\caption{$\varpi=\pi$} \label{2atom}
	\end{subfigure}
	\caption{The evolution of \textit{concurrence} $\mathcal{C}(t)$ vs the scaled time, for the one-photon process ($\kappa=1$), $\theta=\pi/4$, $\omega_{j}=\Omega_{j}=0.1\lambda$, $\gamma_{j}=0$, and $\bar{n}_{j}=10$.}
	\label{concurrence-figures1}
\end{figure}
In figure (\ref{concurrence-figures}), we plot $\mathcal{C}(t)$ versus the scaled time by using the estimated initial value for the latter phase $\phi=\frac{\pi}{2}$ which is chosen due to the existence of the artificial phase jump at this value in the geometric phase in figure(\ref{control-phase}). In figure (\ref{conc3}), we set the atoms initially to be in excited (upper-most) states $\theta=0$, and $\varpi=\pi$, we note that DEM $\leq \ln 2$, which is less than the standard result in models initially prepared in superposition of states. In the next figures, we examine the results of considering superposition of atomic states. In figure (\ref{conc1}), we set the coupling variation parameter $\varpi=0$, and take three various values for the  decay parameter $\gamma$ of the field. We observe that, in the beginning of the interaction between the two atoms and the coherent field $\lambda t \leq 3\pi $, the effect of the decay parameter is approximately not noticed and the concurrence curves are very similar, but at a drastic point of change it differs dramatically as we see that the black curve $\gamma=0$, and then it fluctuates till reaching a stable case of concurrence to be $\geq 0.8$; the red curve $\gamma=(10^{-4}\lambda)$ has a chaotic behavior, as in the beginning. It evolves to give a higher rate of concurrence compared with the absence of decay case (black curve) and after a sufficient time it decreases. The blue curve $\gamma=(10^{-3}\lambda)$ represents the system when concurrence rapidly reaches  the disentangled state (death of entanglement). In figure (\ref{conc2}), we set $\varpi=\pi$, and we note the effect of the oscillation in the matter-field coupling as proposed in the considered model. The effect of that function is clearly noted in the higher case of the decay rate (blue curve) as the interaction has become very weak and fluctuations affect the system evolution. The effect of the time variation function appears as the Frequency Modulation (FM) effect in the radio waves. FM is a method to encode information in a laser field by varying the instantaneous frequency of the coupling between matter and laser. Also, we note that the presence of $\varpi$ or its absence, the system has reached a disentangled state in the same period of scaled time, but the evolution itself changes by the presence of $\varpi$. In figure (\ref{concurrence-figures1}), by taking into consideration the effect of the decay in the atomic energy levels$(\Gamma_{j})$, the concurrence has not been affected.
\begin{figure}[!htb]
	\begin{subfigure}[b]{0.45\columnwidth}
		\centering
		\includegraphics[width=\textwidth,height=3cm]{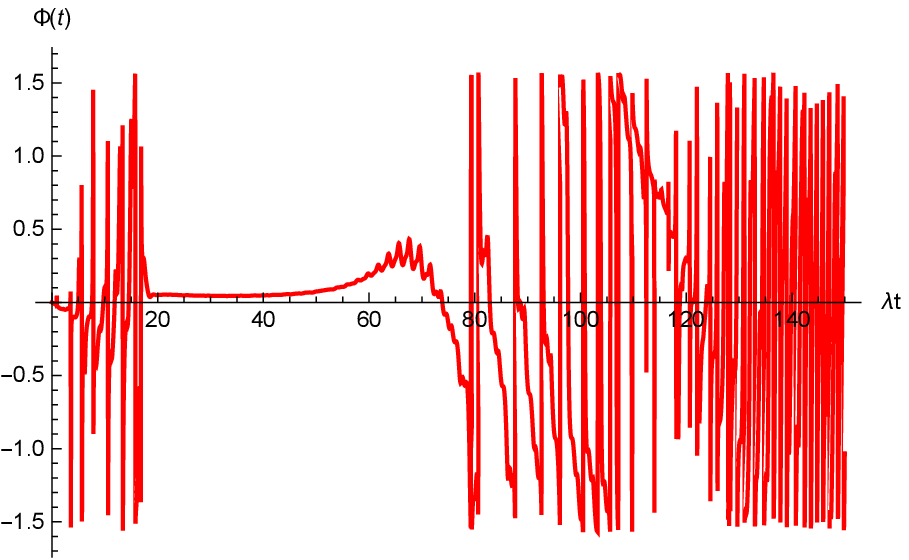}
		\caption{$\varpi=\pi$, and $\gamma=10^{-3}\lambda$}\label{panch1}
	\end{subfigure}
	\quad
	\begin{subfigure}[b]{0.45\columnwidth}
		\centering
		\includegraphics[width=\textwidth,height=3cm]{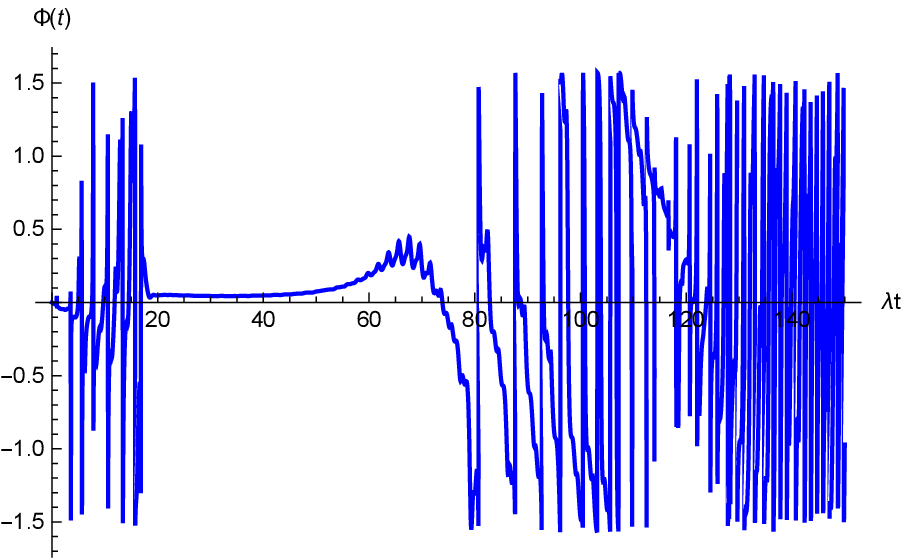}
		\caption{$\varpi=\pi$, and $\Gamma=10^{-4}\lambda$} \label{panch2}
	\end{subfigure}
	\caption{The evolution of  Pancharatnam $\Phi(t)$  vs the scaled time, for the one-photon process ($\kappa=1$), $\theta=\pi/4$, $\omega_{j}=\Omega_{j}=0.1\lambda$, and $\bar{n}_{j}=10$.}
	\label{panch-figures}
\end{figure}

In figure(\ref{panch-figures}), we display the evolution of Pancharatnam phase $\Phi(t)$ vs the scaled time, for various values of the system decay parameters $\gamma$, and $\Gamma$. Both curves approximately exhibit the same behavior and for $\lambda t \leq 6\pi$ the phases exhibits a quick subsequent artificial phase jumps, then take a dominate saturation period till $\lambda t \leq 16 \pi$ which is followed by a slow fluctuation that evolves to start another subsequent artificial phase jumps but less quick than the previous evolution. 
\section{Conclusion}\label{section-conclusion}
\quad The interaction between  atoms with field of the system has been investigated with taking into consideration that cavity leaks energies of both atoms and field while the laser field couples the atom as a cosine wave function of time with a parameter $\varpi$ in the multi-photon process. The RWA has been applied twice to approximate the interaction part of the system. By solving the coupled differential equations resulting by applying the time-dependent Schr\"{o}dinger equation, we get the wave vector and the corresponding eigenenergies. To control the Degree of the Entanglement (DEM) of the system, we determine the initial latter phase by plotting the Pancharatnam phase for three different time points and investigate the concurrence between the two atoms according to the best value of the latter phase. By increasing the effect of  field decay parameter $\gamma$, the concurrence rapidly reaches to the disentangled state (death of entanglement). On the contrary, the atomic decay parameter $\Gamma$ has no effect on the concurrence. The effect of the time-variation function appears as FM effect in the radio waves. FM is used to encode information  between atom and field. The system reaches disentangled state in the same period of scaled time, but the evolution itself changes by the presence of $\varpi$. We note that for various values of the system decay parameters $\gamma$, and $\Gamma$,  the evolution of Pancharatnam phase in both curves approximately exhibits the same behavior.

\bibliographystyle{ieeetr}
\bibliography{bibfile}
\end{document}